\documentclass[iop]{emulateapj}
\usepackage{graphicx}
\usepackage{amsmath}
\usepackage{natbib}

\newcommand{\Msun}{\,\mathrm{M}_{\odot}}
\newcommand{\Lsun}{\,\mathrm{L}_{\odot}}
\newcommand{\feh}{\mathrm{[Fe/H]}}
\newcommand{\kms}{\mathrm{km~s}^{-1}}
\newcommand{\pc}{\,\mathrm{pc}}
\newcommand{\kpc}{\,\mathrm{kpc}}
\newcommand{\Mpc}{\,\mathrm{Mpc}}
\newcommand{\Lmin}{L_{\mathrm{min}}}
\newcommand{\Lmax}{L_{\mathrm{max}}}
\newcommand{\ndof}{n_{\mathrm{dof}}}
\def\cfont{\small}

\begin{document}

\slugcomment{to be submitted to ApJ}
\shortauthors{Harris et al.}
\shorttitle{Globular Cluster Systems in Brightest Cluster Galaxies}

\title{Globular Cluster Systems in Brightest Cluster Galaxies:  A Near-Universal Luminosity Function?}
        
\author{ William E. Harris\altaffilmark{1}, 
         Warren Morningstar\altaffilmark{2}, 
         Oleg Y. Gnedin\altaffilmark{2}, 
         Heather O'Halloran\altaffilmark{1}, 
         John P.~Blakeslee\altaffilmark{3},
         Bradley C.~Whitmore\altaffilmark{4}, 
         Patrick C\^ot\'e\altaffilmark{3},
         Douglas Geisler\altaffilmark{5}, 
         Eric W.~Peng\altaffilmark{6}, 
         Jeremy Bailin\altaffilmark{7}, 
         Barry Rothberg\altaffilmark{8},
         Robert Cockcroft\altaffilmark{1}, 
         and Regina Barber DeGraaff\altaffilmark{9}
}

\altaffiltext{1}{Department of Physics \& Astronomy, McMaster University, Hamilton, ON, Canada; harris@physics.mcmaster.ca, ohallohm@mcmaster.ca, cockcroft@physics.mcmaster.ca}
\altaffiltext{2}{Department of Astronomy, University of Michigan, Ann Arbor, MI 48109; wmorning@umich.edu, ognedin@umich.edu}
\altaffiltext{3}{Herzberg Institute of Astrophysics, National Research Council of Canada, Victoria, BC V9E 2E7, Canada;jblakeslee@nrc-cnrc.gc.ca, patrick.cote@nrc-cnrc.gc.ca}
\altaffiltext{4}{Space Telescope Science Institute, 3700 San Martin Drive, Baltimore MD 21218, USA; whitmore@stsci.edu}
\altaffiltext{5}{Departamento de Astronomi\'a, Universidad de Concepci\'on, Casilla 160-C, Concepci\'on, Chile; dgeisler@astroudec.cl}
\altaffiltext{6}{Department of Astronomy, Peking University, Beijing 100871, China; peng@bac.pku.edu.cn}
\altaffiltext{7}{Department of Physics and Astronomy, University of Alabama, Box 870324, Tuscaloosa, AL 35487-0324, USA; jbailin@ua.edu}
\altaffiltext{8}{LBT Observatory, University of Arizona, 933 N.Cherry Ave, Tucson AZ 85721, USA; dr.barry.rothberg@gmail.com}
\altaffiltext{9}{Department of Physics and Astronomy, Western Washington University, Bellingham WA 98225, USA; Regina.BarberDeGraaff@wwu.edu}

\date{\today}

\begin{abstract}
We present the first results from our HST Brightest Cluster Galaxy (BCG) survey of seven 
central supergiant cluster galaxies and their globular cluster (GC) systems.  We measure a 
total of 48000 GCs in all seven galaxies, representing the largest 
single GC database. We find that a log-normal shape accurately matches the observed
the luminosity function (LF) of the GCs down to
the GCLF turnover point, which is near our photometric limit.  In addition, the LF 
has a virtually identical shape in all seven galaxies.  
Our data underscore the similarity in the formation mechanism of massive star clusters in 
diverse galactic environments.
At the highest luminosities (log $L \gtrsim 10^7 L_{\odot}$) we find small
numbers of ``superluminous'' objects in five of the galaxies; their luminosity and color
ranges are at least partly consistent with those of UCDs (Ultra-Compact Dwarfs).
Lastly, we find preliminary evidence that in the outer halo ($R \gtrsim 20$ kpc),
the LF turnover point shows a weak dependence on projected distance, scaling as $L_0 \sim R^{-0.2}$,
while the LF dispersion remains nearly constant.
\end{abstract}

\keywords{galaxies: formation --- galaxies: star clusters --- 
  globular clusters: general}

\section{Introduction}

The luminosity function -- the number of objects per unit magnitude or unit luminosity -- is one 
of the most fundamental properties of globular cluster (GC) systems.  The globular cluster luminosity function (GCLF) 
is the observational 
surrogate for the more astrophysically relevant GC \emph{mass function} (GCMF); in turn, what we see at the present day 
is the product of the mass distribution of the clusters at their time of formation and the
dynamical evolution of the GCs within their host galaxy over a Hubble time.
In the Milky Way, 
either the GCLF or GCMF were long ago found to have a unimodal and nearly symmetric shape, 
often fit by a log-normal function \citep[e.g.,][2010 edition, for the latest homogeneous collection of data]{harris96}.

Because the formation histories and tidal fields of galaxies can differ strongly from one another, 
it continues to be somewhat surprising that in galaxies of all morphological types and sizes, the GCLF 
has a nearly identical characteristic log-normal shape \citep[e.g.,][]{harris2001,brodie_strader06,jordan2007}.  
The ``turnover'' luminosity $L_0$ where the LF peaks increases weakly 
with galaxy mass \citep{jordan2007,villegas2010}. However, to first order it remains 
true that the turnover is so nearly universal that it can be used as a 
standard candle for extragalactic distance determination to a typical accuracy of a quarter-magnitude;
if attention is paid to the second-order dependence of $L_0$ on galaxy size \citep{villegas2010}
then the accuracy can approach $\pm 0.15$ magnitude particularly for large galaxies \citep[see][for a recent review]{rejkuba2012}.

Although in rough outline the present-day form of the GCLF is understood, quantitative details are still uncertain.
Since the stellar mass-to-light ratio does not vary much with luminosity for the old stellar populations of GCs, 
a universal GCLF translates directly to a universal GC mass function.  But a peaked distribution differs radically 
from the observed mass function of star clusters in nearby interacting and starbursting galaxies 
\citep[e.g.,][]{zhang_fall99, gieles09, larsen09, whitmore_etal2010, chandar_etal11}, where extremely young clusters can be seen over a wide range
of masses.  This initial mass function is typically consistent with a single power law, 
$dN/dM \propto M^{-\beta}$ with $\beta \approx 1.8-2.2$ over a range of cluster mass from 
$10^{5}\Msun \lesssim M \lesssim 10^{7}\Msun$. Models of the dynamical evolution and disruption of GCs 
\citep[e.g.][among many others]{fall_zhang2001,baumgardt_makino2003,gieles_baumgardt08,mclaughlin_fall2008,kruijssen_etal2011} 
show that the fractional mass loss is strongly 
dependent on cluster mass; lower-mass clusters are preferentially 
dissolved over a Hubble time by internal evaporation coupled to the external tidal 
field, tidal shocking, and dynamical friction.  
These mechanisms seem capable of transforming an initial power-law MF 
into a log-normal one dominated by the highest-mass clusters, at least over the mass range that
covers most clusters.  Above a few $10^5 M_{\odot}$, however, mass loss rates are relatively slow
and such massive clusters should more nearly preserve the initial MF shape.

Current attempts to incorporate the formation and evolution of GCs in the hierarchical framework of galaxy formation
have had some reasonable initial success
at reproducing not only the observed GCMF but also the cluster  metallicity distribution in the 
Milky Way \citep{muratov_gnedin10, griffen2010} and in giant early-type galaxies \citep{li_gnedin14}. 
GCs are simply the most massive star clusters, and like all star clusters, they are likely to have formed as
initial high-mass gaseous clumps of typically $\sim 1$ pc scale size embedded within much more extended giant
molecular clouds (GMCs) \citep[e.g.][]{harris_pudritz1994,burgarella2001,bromm_clarke2002,beasley2002,
shapiro_etal2010,elmegreen2012}.
Detailed cosmological simulations of the formation of high-redshift galaxies predict the existence of giant 
molecular clouds (GMCs) with the right large masses, sizes, and low 
metallicities within which embedded proto-GCs can form and eventually
evolve into the extremely dense, massive GCs of the present day \citep{kravtsov_gnedin05}.  
A simple sub-grid model for the creation of star clusters at gas density above a fixed threshold ($10^4\Msun\pc^{-3}$) 
results in a GCMF consistent with the $\beta \approx 2$ power law.  Alternatively, the mass function could 
be described by a curved log-normal shape but with very different parameters than the present-day GCMF,  
motivated by the probability distribution function of clumps in a turbulent interstellar medium. 
Both functional forms are possible when fit over a relatively narrow range of cluster mass,
but differences between them may occur at the highest cluster masses.  
Therefore, extending the measurement of the GCLF (and GCMF) to the most massive clusters offers the
potential to test models of GC formation based on supersonic turbulence in GMCs.

Although the theoretical modelling mentioned above is more advanced than in earlier years, the numerical problem 
is essentially that galaxy-formation simulations must cover scales of $\sim 10^{12} M_{\odot}$ and $\sim 100$ kpc, while full resolution of 
proto-GCs requires resolving scales \emph{at least} $10^7$ times smaller in both quantities, an extremely steep
challenge to accomplish simultaneously \citep[see][for discussion]{harris2010}.  Modelling work directed at bridging this gap
is ongoing \citep[e.g.][]{bournaud2008, mashchenko2008, howard2014} but a complete theory of GC formation is not yet in hand.
It is, however, globally important because a significant fraction of star formation and stellar feedback should happen in these 
densest, most massive clumps.

The highest-mass GCs are especially interesting as well because of their potential connections to the populations of
ultra-compact dwarf galaxies (UCDs) that are typically found in rich clusters of galaxies.  In structural parameter planes
of effective radius, luminosity, or surface brightness the UCDs bridge the traditional gap between the GC sequence and the 
dwarf ellipticals \citep[e.g.][among many others]{hasegan2005,misgeld_etal11,brodie2011,penny2014}.  
UCDs may be especially luminous GCs, remnant nuclei of dwarfs, remnants 
of multiple mergers of GCs, or genuine small-scale dwarfs.  Because they are relatively rare and hard to find, their possible
connections with GCs remain uncertain.

GCs with luminosities in the range $L ~\sim 10^4 - 10^6 L_{\odot}$ (i.e. the two orders of magnitude centered on the turnover
luminosity) are by far the most numerous and so the shape of the GCLF is observationally well established in
this middle range.  The high-mass end (roughly speaking, 
objects like $\omega$ Centauri and above, with masses up to $10^7 M_{\odot}$ and beyond) is much less well surveyed.  
Because they are rare, very large statistical samples of GCs are needed to fill in this high-mass region.

The most effective way to accumulate the largest possible samples of GCs is through observation of the most 
luminous elliptical galaxies -- the cD or BCG (Brightest Cluster Galaxy) systems that reside at the centers 
of rich galaxy clusters.  Because of the empirical fact that the GCLF dispersion increases systematically with galaxy size
\citep{villegas2010},
simply stacking large numbers of smaller galaxies to gain a high$-n$ total sample is not equivalent to using a single supergiant 
galaxy with the same total $n(GC)$ \citep[see, e.g.,][for a specific example constructed from the Virgo and 
Fornax surveys]{mieske2010}.
In the present paper, we give a brief overview of a new Hubble Space Telescope (HST) survey of GC systems in seven 
BCGs and discuss our first findings for the observed GCLF.  In following papers of this series, we will present 
the complete database along with more 
thorough discussions of the color-magnitude diagrams, the GC metallicity distributions, and related issues relevant 
to the formation histories of our target galaxies.

\begin{table*}[t]
\begin{center}
\caption{\sc Basic Parameters for BCGs in the Survey}
\label{tab:basicdata}
\begin{tabular}{llllrccl}
\tableline\tableline\\
\multicolumn{1}{l}{Galaxy} &
\multicolumn{1}{l}{Cluster} &
\multicolumn{1}{l}{RA} &
\multicolumn{1}{l}{Dec} &
\multicolumn{1}{l}{$cz \, (\kms)$} &
\multicolumn{1}{l}{$(m-M)_I$} &
\multicolumn{1}{l}{$A_I$} &
\multicolumn{1}{l}{$M_V^T$}
\\[2mm] \tableline\\
NGC 7720    & A2634 & 23:38:29 & $+$27:01:53 &  8714 & 35.61 & 0.107 & -23.35 \\
NGC 6166    & A2199 & 16:28:38 & $+$39:33:06 &  9125 & 35.60 & 0.017 & -23.7  \\
UGC 9799    & A2052 & 15:16:44 & $+$07:01:18 & 10500 & 35.95 & 0.056 & -23.1  \\
UGC 10143   & A2147 & 16:02:17 & $+$15:58:29 & 10741 & 35.99 & 0.047 & -23.0  \\
ESO509-G008 & A1736 & 13:26:44 & $-$27:26:22 & 10848 & 36.06 & 0.079 & -23.35 \\
ESO383-G076 & A3571 & 13:47:28 & $-$32:51:54 & 11832 & 36.24 & 0.082 & -24.1  \\
ESO444-G046 & A3558 & 13:27:57 & $-$31:29:44 & 14345 & 36.65 & 0.075 & -23.8
\\[2mm] \tableline
\end{tabular}
\end{center}
\vspace{0.1cm}
\end{table*}

\begin{table}
\begin{center}
\caption{\sc Parameters for the Imaging}
\label{tab:observations}
\begin{tabular}{lrlrll}
\tableline\tableline\\
\multicolumn{1}{l}{Galaxy} &
\multicolumn{2}{c}{F475W} &
\multicolumn{2}{c}{F814W} &
\multicolumn{1}{l}{$R$ Range} \\[0.5mm]
\multicolumn{1}{l}{} &
\multicolumn{1}{r}{time (s)} &
\multicolumn{1}{l}{lim mag} &
\multicolumn{1}{r}{time (s)} &
\multicolumn{1}{l}{lim mag} &
\multicolumn{1}{l}{(kpc)}
\\[2mm] \tableline\\
NGC 7720    &  5282 & 28.35 &  5278 & 27.95 & 12-110 \\
NGC 6166    &  5370 & 28.5  &  4885 & 27.9  & 12-100 \\
UGC 9799    &  7977 & 28.45 &  5253 & 27.77 & 15-105 \\ 
UGC 10143   & 10726 & 28.7  &  5262 & 27.55 & 15-130 \\
ESO509-G008 & 10758 & 28.65 & 18567 & 28.6  & 15-120 \\
ESO383-G076 & 10830 & 28.7  & 21081 & 28.5  & 16-130 \\
ESO444-G046 & 21660 & 29.25 & 34210 & 28.6  & 20-170
\\[2mm] \tableline
\end{tabular}
\end{center}
\vspace{0.1cm}
\end{table}

\section{Brightest Cluster Galaxy Survey}
  \label{sec:survey}

BCGs contain the most populous GC systems, but they are also a rare type of galaxy and so we must look much 
further afield than the closest galaxy clusters to find more than just the few that have been studied 
to date \citep{bassino2006,harris09a,harris09b,wehner_etal08,peng_etal11}.  
The paradigm of a GC system in a BCG-type galaxy is that in M87, which holds 13,000 clusters \citep[e.g.][]{harris09b}.  
However, larger systems are known \citep[see][]{harris_etal13}, such as the Coma central cD NGC 4874 that may hold as 
many as 30,000 GCs \citep{peng_etal11}.  
An even larger system may be that around Abell 1689 
at $cz = 54000$ km s$^{-1}$ \citep{alamo-martinez_etal2013}.
Many more BCG targets of the size of NGC 4874 or even larger have GC systems that
are within reach of HST.  

\subsection{Observational Material}

In this paper we present the first results of photometry for GCs in
seven BCG galaxies, all of which are examples of the largest galaxies that exist in the present-day universe.  
Each of these was already known from previous deep ground-based imaging to contain
a rich GC population \citep{harris_etal1995,bridges_etal1996,blakeslee_etal1997}.  Our targets, listed in Table~\ref{tab:basicdata} in order of increasing
distance, were imaged during program GO-12238 (Harris, PI) with the exception of the $F814W$ 
exposures for the three ESO galaxies, which came from program GO-10429 (Blakeslee, PI).  
The Table lists in successive columns the
galaxy ID; the Abell cluster for which it is the central BCG; coordinates; redshift of the galaxy $cz$ normalized to the
CBR reference frame; apparent distance modulus; foreground extinction (from NED on the Landolt
scale); and $V-$band luminosity, where $V_T$ is taken from NED. All the foreground extinctions
are small and have no effect on the following analysis.
Throughout the following discussion we adopt a distance scale of $H_0 = 70\,\kms\Mpc^{-1}$.
The uncertainties in the conversion to the CBR frame, and possible motion
of the BCG relative to the Abell cluster center, may make the true $cz$ differ by up
to $\sim 400$ km s$^{-1}$, or $0.05 - 0.10$ in distance modulus depending on
distance.

The primary camera for all targets was ACS/WFC, with WFC3 being used in parallel;
exposures in both cameras were through the F475W and F814W filters.  In
choosing these targets, to save orbits we took advantage of the long exposures in
F814W of the three ESO galaxies that were already in the HST Archive.
The luminosities of these giant galaxies are comparable with or even
higher than the two brightest BCGs whose GC systems have previously
been studied to similar depth, namely NGC 4874 in Coma and NGC 4696 in Centaurus
\citep{harris09a}.

In Table~\ref{tab:observations}, we summarize the raw exposure data
for the ACS/WFC pointings, including the total exposure times and the
limiting magnitudes that resulted.  Two or more orbits were used for
each filter, subdivided into sequences of half-orbit exposures.
The total exposures were designed to reach a limit in absolute magnitude
close to the expected GCLF turnover point at $M_{V,0} \simeq -7.3, M_{I,0} \simeq -8.4$
and thus to secure the largest possible GC sample sizes within the limits
of the program.

Preprocessing consisted of CTE correction on the raw images 
\citep{anderson_bedin2010}, and then
use of the standard Multidrizzle package to generate a combined image
in each filter that kept the native pixel scale of $0.05\arcsec$.

\subsection{Photometric Reductions}

Photometry was carried out with the standard tools in SourceExtractor
\citep{bertin_arnouts96} and DAOPHOT \citep{stetson87} in its IRAF
implementation, including aperture photometry (\emph{phot}) and
then point-spread-function (PSF) fitting through \emph{allstar}.
All PSFs were empirically generated from stars on the target fields,
with anywhere from 60 to 160 individual stars on each frame.  Finding
bright, isolated candidate PSF stars was easily done, since our
targets are all moderately high-latitude fields that are completely
uncrowded in any absolute sense.

Each target galaxy was found, as expected, to be surrounded by many
thousands of clusters.  Virtually all of the individual GCs appear
starlike: since our target systems are in the distance range $125 -
205$ Mpc, a typical GC effective radius of $\sim 3\pc$ corresponds to
an angular size in the range $0.003\arcsec-0.005\arcsec$, far below the
$0.1\arcsec$ resolution of the telescope and safely in the
``unresolved'' category according to the criteria in
\citet{harris09a}.  The direct advantage for our photometry is that
the GCs are therefore easily distinguished from the faint, nonstellar
background galaxies that constitute the main source of sample
contamination.

In the last column of Table~\ref{tab:observations}, we also give the 
radial range $R(min) - R(max)$ from galaxy center covered by the ACS field.  This 
$R(max)$ is determined both by the distance to the galaxy and its placement relative
to the center of the ACS/WFC field of view.  The innermost radii sampled for any of the galaxies
are typically $\sim 20$ arcsec, corresponding to $R(min) \sim 10-20$ kpc depending on
distance; at smaller radii the object detection and photometry are more severely 
limited by the higher surface brightness of the central galaxy.  These numbers show that the bulk of our
GC sample is drawn from the mid- to outer-halo regions of these supergiant
galaxies.

The calibration of our data is in the VEGAMAG system, and uses the
most recent filter-based zeropoints given on the HST webpages to
define the $F475W$ and $F814W$ natural magnitudes.  For purposes of
the present discussion, and ease of comparison with previous work
\citep[e.g.][]{harris_etal06, harris09a}, we use conventional $(B,I)$
magnitudes converted from $(F475W, F814W)$ defined by the transformations in \citet{saha_etal11}.
Using the reddening corrections for each galaxy from the NED database (listed in Table~\ref{tab:basicdata}), 
we convert the preliminary color-magnitude distributions to absolute
magnitude $M_I$ and intrinsic color $(B-I)_0$.  

\begin{figure*}[ht]
\begin{center}
\includegraphics[width=0.8\hsize]{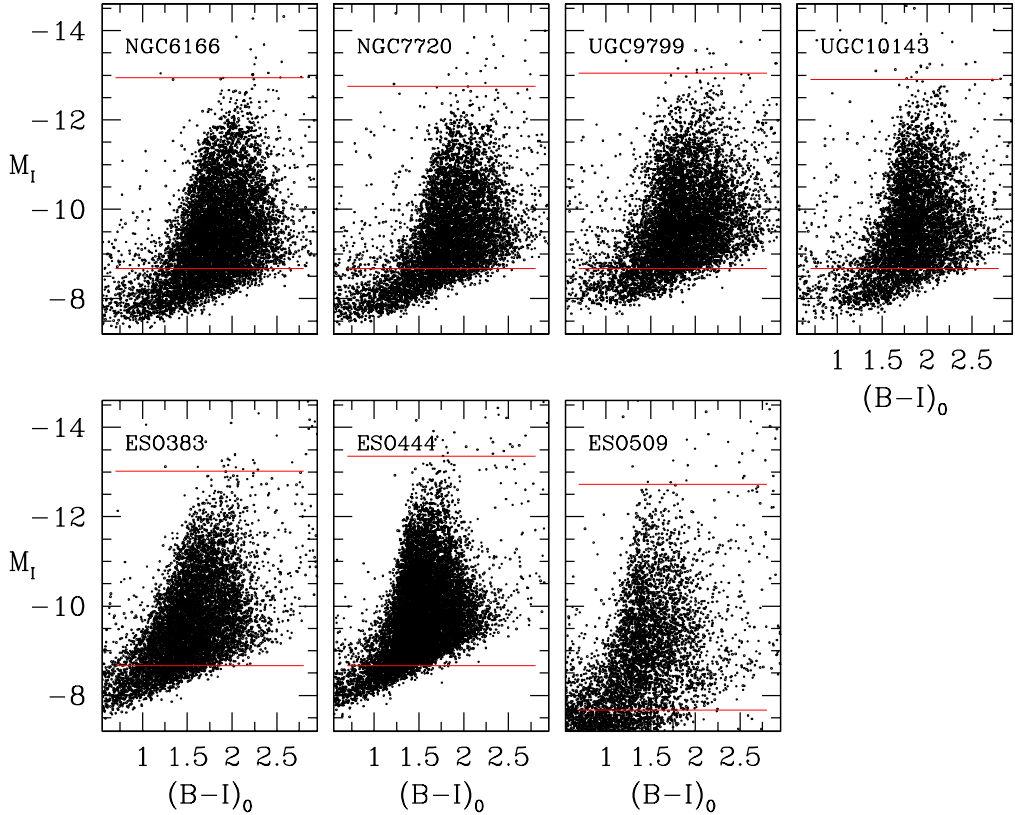}
\end{center}
  \vspace{-2.6cm}
\caption{\cfont Color-magnitude diagrams for the seven
  BCGs in our sample.  The filter-based magnitudes ($F475W, F814W$)
  have been converted to ($B,I$) as described in the text, 
  and intrinsic colors and absolute magnitudes
  are calculated from the distance moduli and reddenings in
  Table~\ref{tab:basicdata}. For comparison, the horizontal lines in each panel show
  the values of $L_{min}$ and $L_{max}$ within which the Gaussian LF fitting was done.
  The ``superluminous'' objects discussed in Section 4 of the text are ones
  brighter than $L_{max}$.
  }
  \vspace{0.2cm}
  \label{fig:mdf7}
\end{figure*}

Figure~\ref{fig:mdf7} shows the color-magnitude diagrams for all seven galaxies.  
The huge GC populations around each one are evident, and to first order the distributions in
luminosity and color are similar.  To quantify the limits of the
photometry, we carried out an extensive series of artificial-star
tests on every field through \emph{daophot/addstar}.  The limiting
magnitudes quoted in Table~\ref{tab:observations} (not de-reddened, and in
the natural filter-based VEGAMAG scale) for each field and
filter are the levels at which the artificial-star tests show that the
completeness of detection falls to 50\%.  At levels $\gtrsim 0.5$~mag
brighter, the completeness approaches 100\%.
We find that within the radial range $R \gtrsim 20$ kpc that contains
the vast majority of our measured GCs, 
the surface brightness of the BCG itself has fallen to low
enough levels that the limiting magnitudes do not depend noticeably on $R$.
As noted above, crowding effects are also negligible in this same radial range.
It is worth noting that this situation is in strong contrast to imaging of much 
nearer galaxies such as the 
Virgo members at $d \simeq 16$ Mpc \cite[e.g.][]{peng_etal08}, where the ACS field of view
covers a \emph{maximum} radius of only $R \sim 10$ kpc and the surface brightness gradient of
the central galaxy across the detector is more of a concern.

\begin{figure*}[ht]
\vspace{-0.1cm}
\begin{center}
\includegraphics[width=0.5\hsize]{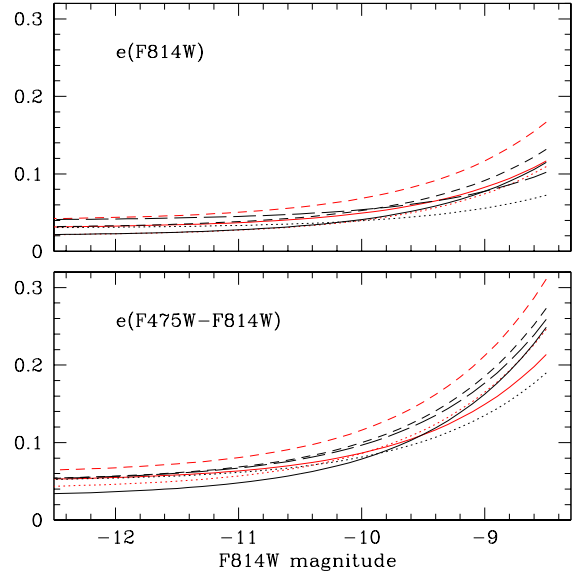}
\end{center}
\vspace{-0.3cm}
\caption{\cfont Measurement uncertainties in magnitude $F814W$ and color $(F475W-F814W)$,
plotted as a function of absolute $I-$band magnitude.
\emph{Solid red line:} NGC 6166.
\emph{Dotted red line:} NGC 7720.
\emph{Dashed red line:} UGC 9799.
\emph{Solid black line:} UGC 10143.
\emph{Dotted black line:} ESO509.  
\emph{Dashed black line:} ESO383.  
\emph{Dot-dashed black line:} ESO444. } 
\vspace{0.2cm}
  \label{fig:photerr}
\end{figure*}

In Figure~\ref{fig:photerr} we show the trend of photometric measurement uncertainties
	for each galaxy, plotted as a function of absolute magnitude $M_{F814W} \simeq M_I$.  The exposure times were planned to yield 
fairly uniform limits in absolute magnitude, so these curves are similar for all galaxies
in the sample.  As will be seen below, the
uncertainties in the $F814W$ magnitudes are $\lesssim 0.1$ mag for all levels
brighter than the GCLF turnover, implying that the 
$\simeq 1.3-$mag intrinsic width of the GCLF is 
negligibly broadened by photometric scatter. 

Any field contamination of the GC samples is due to sparsely distributed
foreground stars and a few very small, faint unresolved background galaxies.  To
gauge the level of contamination we used the offset WFC3 fields, placed on
relatively galaxy-free locations in the outskirts of the galaxy cluster.
After the same photometric procedure, rejection of nonstellar objects, and rejection of objects 
with extreme colors (see below), we found
that the residual contamination (and thus its effects on the measured
GC luminosity functions) was negligible at any level brighter than the
limiting magnitudes of the survey.  

The CMDs generally show evidence for broad bimodal or multimodal distributions in the GC colors.
More extensive discussion will be
given in following papers that concentrate on the color-magnitude and spatial distributions.

\begin{figure*}[ht]
\vspace{-0.6cm}
\begin{center}
\includegraphics[width=0.9\hsize]{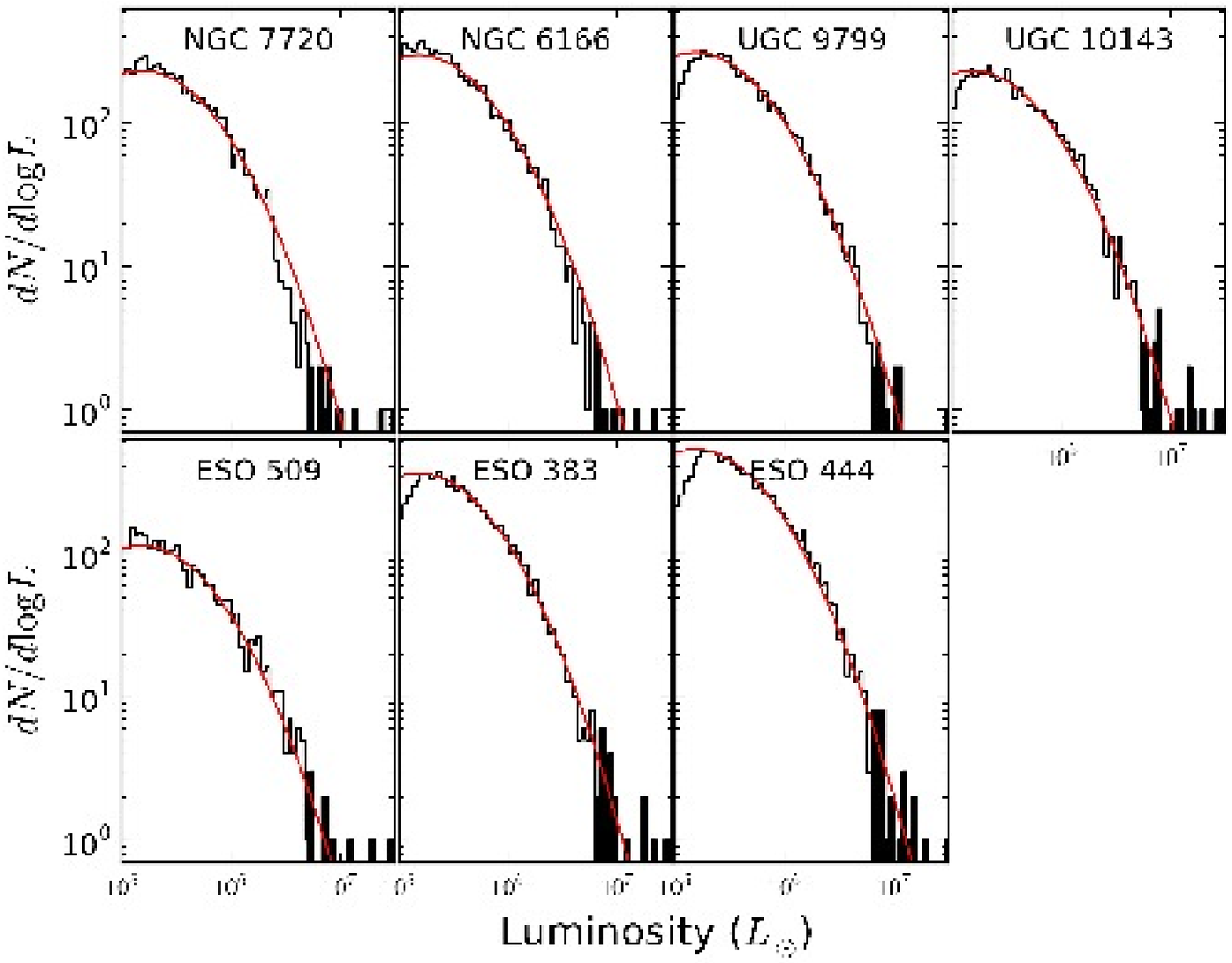}
\end{center}
\vspace{-0.3cm}
\caption{\cfont Luminosity function of globular cluster systems in the seven target galaxies (black histogram).  
Solid lines overplot the best log-normal fit for the combined global sample.  
Shaded parts of the histograms show super-luminous objects, defined by Equation~(\ref{eq:lmax}).
Note that in this graph, a simple power-law form for $N(L)$ would be a straight line.}
\vspace{0.2cm}
  \label{fig:lf_all}
\end{figure*}

\section{The Luminosity Function}

\subsection{GCLF fitting}

To reduce contamination of our sample, we select only objects within the 
dereddened color range $1.0 < (B-I)_0 < 2.5$. A cutoff at the blue end eliminates any objects bluer than the 
minimum metallicity known for GCs ($\feh < -2.5$) \citep[see][]{harris09a}.
Translation from $(B-I)_0$ to metallicity 
[Fe/H] here assumes the linear relation $(B-I)_0 = 2.158 + 0.375$[Fe/H] \citep{harris_etal06}.
The red-end cut is based on empirical increase of contaminants in the parallel background fields, which are
more frequent at colors redder than the GC sequences.
In the case of ESO444, we have also cross-correlated our objects with the catalog 
of nucleated dwarf galaxies identified by John Blakeslee in program GO-10429. 
We found only 4 objects in common and eliminated those.  The $0.02''$ angular resolution limit 
above which we could reliably identify nonstellar objects \citep[see][]{harris09a}
corresponds to a linear resolution of $\sim 20$ pc 
at the average distance of our BCG targets.  This limit is not low enough to distinguish many or
most UCDs from luminous GCs, so our sample is likely to include some UCDs at the high$-L$ end (see the discussion 
in Section 4 below).

Figure~\ref{fig:lf_all} shows the resulting GCLFs for each of the seven BCGs. 
At the lowest plotted luminosity, $L_{I} = 10^{5}\Lsun$ (which is closely similar to
the fiducial GCLF turnover luminosity), three systems are 
complete to at least 50\% at that level and the others are essentially complete to $\sim 100\%$.

We fit a log-normal (Gaussian in $\log{L}$) distribution to the GCLF in the I-band
\begin{equation} 
  {dN \over d\log{L}} = N_{0}\, 
    \exp{\left[ -{(\log{L}-\log{L_0})^{2} \over 2\sigma_{L}^{2}}\right]},
  \label{eq:gaussianfit}
\end{equation}
for clusters within a chosen range $\Lmin < L_I < \Lmax$ described below.
(NB:  by $\log$ we denote logarithm base-10.)  To convert $M_I$ to $L_I$ we adopt $M_I(\odot) = 4.08$.
The two free parameters are the turnover (peak) luminosity $L_0$ and the Gaussian dispersion $\sigma_{L}$, while $N_0$ 
is constrained by the total number of clusters. 

We bin the data evenly spaced in $\log{L}$ and for consistency, we use the same bin size for all seven 
systems in the survey. We have experimented with varying the bin size from 0.01~dex to 
0.1~dex and calculated the average $\chi^2$ per number of degrees of freedom ($\ndof$) of 
individual galaxy fits.  It is defined as
\begin{equation}
  \chi^2 = \sum_i {(N_{i,\rm obs} - N_{\rm exp}(L_i | L_{0},\sigma_{L},N_{0}))^2 \over (\Delta N_{i,\rm obs})^2},
\end{equation}
where $N_{i,\rm obs}$ is the observed and completeness-corrected number of clusters in bin~$i$, $N_{\rm exp}$ is the expected 
number from the fitting function above, and $\Delta N_{i,\rm obs} = N_{i,\rm obs}^{1/2}$ is 
the Poisson counting uncertainty. The number of degrees of freedom is the number of bins 
minus 3, accounting for $L_0$, $\sigma_{L}$, and $N_0$.
The minimum $\chi^2/\ndof$ 
occurs at $\delta\log{L} = 0.02$~dex, which we adopt for our analysis. This 
choice also results in a statistically optimum number of bins $\approx N_{\rm tot}^{1/2}$.

Limiting the range of luminosity for constraining the fit parameters is necessary because 
at low $L$ the cluster counts are incomplete in both $B$ and $I$, and at high $L$ the log-normal function does not 
account for any superluminous clusters that form an extended tail to the LF (see Section 4 below).  
We have varied $\log{\Lmin}$ (in Solar units) from 
5.0 to 5.3 in steps of 0.1, and calculated $\chi^2/\ndof$ for each system.  Values of 5.1 and 5.2 gave 
similar fits for most galaxies, and therefore, we adopt $\log{\Lmin}=5.1$ to include more clusters in our analysis. 
This limit is also conservatively brighter (by about 1 magnitude) than the $I-$band completeness limit of the photometry, 
but is near the $B-$band limit for the reddest GCs.  
For one system (ESO509), we 
found that lowering $\log{\Lmin}$ to 4.7 gave the most robust fit.

The upper $\Lmax$ to the fitted range  is determined by an iterative procedure.  
First, we set $\Lmax \rightarrow \infty$ and find the best-fitting log-normal function for all clusters 
above $\Lmin$.  Given this fit, we define $\Lmax$ as the limit above which the integrated GCLF predicts only 10 clusters:
\begin{equation}
  \int_{\Lmax}^{\infty} \frac{dN_{\rm exp}}{dL} \, dL = 10.
  \label{eq:lmax}
\end{equation}
The number 10 was chosen based on empirical comparison with the high-$L$ tail of the observed GCLF
so that small-number statistics in the uppermost $L-$bins would not unduly influence the fit.
Integration of Equation~(\ref{eq:gaussianfit}) gives an explicit non-linear relation for $\Lmax$:
\begin{equation}
  N_{0}\, \sigma_{L}\, \sqrt{\frac{\pi}{2}}\, \mathrm{erfc}{\left(\frac{\log{\Lmax}-\log{L_0}}{\sqrt{2}\, \sigma_{L}}\right)} = 10.
  \label{eq:integral_solution}
\end{equation}
For a given combination ($L_0$, $\sigma_{L}$, $N_{0}$), we solve this equation numerically and use the 
new value of $\Lmax$ to repeat the GCLF fit. This process is iterated until $\Lmax$ varies 
by less than $1\%$. The values of $\Lmax$ are determined for each system individually.
Clearly $\Lmax$ is not a fixed boundary because it will be higher for galaxies with larger total populations $N_0$.
However, defining it this way ensures that we will be fitting the fiducial Gaussian function over the luminosity
range where the number of clusters per bin is satisfactorily large.
In Fig.~\ref{fig:mdf7} we show the adopted $L_{min}$ and $L_{max}$ levels 
for each galaxy.

\begin{table*}
\begin{center}
\caption{\sc Luminosity Function Parameters}
\label{tab:par}
\begin{tabular}{lrrccccccrrl}
\tableline\tableline\\
\multicolumn{1}{l}{Galaxy} &
\multicolumn{1}{l}{$N(L_{I}>L_0)$} &
\multicolumn{1}{l}{$N_{SL}$} &
\multicolumn{1}{l}{$\log{\Lmax}$} &
\multicolumn{1}{l}{$\log{L_{I,\mathrm{lim}}}$} &
\multicolumn{1}{l}{$\log{L_0}$} &
\multicolumn{1}{l}{$\Delta\log{L_0}$} &
\multicolumn{1}{l}{$\sigma_{L}$} &
\multicolumn{1}{l}{$\Delta\sigma_{L}$} &
\multicolumn{1}{l}{$\chi^2/(\ndof)$} &
\multicolumn{1}{l}{$\chi^2_{\rm g}/(\ndof)$} &
\multicolumn{1}{l}{$P_{\rm KS,g}$}
\\[2mm] \tableline\\
NGC 7720    & 4372 & 16 & 6.73 & 4.7 & 5.15 & 0.04 & 0.52 & 0.02 & 105/(79) & 121/(74) & 0.011 \\
NGC 6166    & 6043 & 10 & 6.81 & 4.7 & 5.09 & 0.04 & 0.55 & 0.02 &  96/(83) & 119/(78) & 0.0007 \\
UGC 9799    & 4765 & 18 & 6.85 & 4.9 & 5.26 & 0.02 & 0.51 & 0.01 &  74/(85) &  69/(80) & 0.91 \\
UGC 10143   & 3674 & 24 & 6.79 & 5.0 & 5.25 & 0.03 & 0.52 & 0.02 &  83/(82) &  77/(77) & 0.29 \\
ESO509-G008 & 2857 & 18 & 6.72 & 4.6 & 5.07 & 0.03 & 0.58 & 0.02 & 116/(99) &  97/(75) & 0.86 \\
ESO383-G076 & 5217 & 29 & 6.84 & 4.7 & 5.30 & 0.02 & 0.49 & 0.01 &  93/(84) &  74/(79) & 0.72 \\
ESO444-G046 & 7083 & 24 & 6.97 & 4.8 & 5.34 & 0.02 & 0.49 & 0.01 & 165/(91) & 115/(86) & 0.023 \\[1mm]
All (global fit) & & & 7.14 & & 5.24 & 0.01 & 0.52 & 0.01 & 204/(100) & & 
\\[2mm] \tableline
\end{tabular}
\end{center}
{\sc Notes:} (a) $N_{SL}$ is the observed number of ``superluminous'' objects with $L_{I}>\Lmax$.  
(b) $\Lmax$ is the luminosity above which the log-normal fit predicts 10 clusters (see text). 
(c) $L_{I,\mathrm{lim}}$ is the 50\% completeness limit.  
(d) $L_0$, $\sigma_{L}$, and $\chi^2$ of the log-normal fit are calculated in the luminosity range 
from $10^{5.1}\Lsun$ to $\Lmax$, with the individual value of $\Lmax$ for each galaxy. 
(e) $\chi^2_{\rm g}$ and $P_{\rm KS,g}$ of the global fit are calculated over the range $L_0 - \Lmax$.\\
\vspace{0.4cm}
\end{table*}

The best-fit parameters $(L_0, \sigma)$ for individual fits to each GC system are listed in 
Table~\ref{tab:par}. The 68\%-confidence errorbars, $\Delta\log{L_0}$ and $\Delta\sigma_{L}$, are 
calculated at $\Delta\chi^2 = 1$. The table also lists the de-reddened 50\% completeness limits 
$L_{I,\rm lim}$, which are fainter than our adopted lower cut for all systems. The fraction of 
missing (undetected) clusters between $\Lmin$ and the peak of the GCLF, $L_0$, relative to that expected 
from the fit, ranges between 18\% and 29\%.  We conclude that our samples are complete enough 
in the chosen luminosity range for accurate fitting to be carried out.

The observed number of clusters above the peak (shown in the second column in the table) can be roughly doubled to
estimate the total expected population of the GC system, assuming the log-normal GCLF is symmetric.
The total number of clusters per galaxy therefore ranges from about 6,000 to over 14,000. 
The currently detected number of GCs above the 50\% completeness limit for all 
seven systems is 47,910 and the number of clusters above $L_{I} = 10^5\Lsun$ is 43,783.  
This sample represents the largest single dataset of GCs in the literature, an order of magnitude larger,
for example, than the total number of GCs brighter than the turnover in the Virgo survey \citep{jordan2007}.

Table~\ref{tab:par} also lists the $\chi^{2}$ of the Gaussian fit and the number of 
degrees of freedom, $\ndof$.  For each system, the ratio of the two is between 
0.87 and 1.8, which indicates that the log-normal is an appropriate fitting function for these largest GC systems.

The values of peak luminosity and the width for the seven systems 
are remarkably similar to each other; unweighted direct means of the two quantities are
$\langle {\rm log} L_0 \rangle = 5.21 \pm 0.04$ and $\langle \sigma_L \rangle = 0.52 \pm 0.01$. 
The observed rms scatter in log $L_0$ is $\pm0.10$ dex; the expected 
$\sim 0.03-$dex scatter due simply to distance uncertainties (see Section 2 above) does not
contribute significantly to that.
In short, we have no evidence that the seven LFs differ systematically from each other in
any major way.

To further examine this universality, we combine the clusters from all seven galaxies and 
perform a global log-normal fit to this combined sample. For consistency, we keep the same 
lower limit $\Lmin$, but calculate $\Lmax$ as described above. The last row of 
Table~\ref{tab:par} lists the parameters of this global fit: 
$\log{L_{0,\rm g}} = 5.24$, $\sigma_{L,\rm g} = 0.52$~dex. 
As expected, they fall in the middle of the 
range for the individual systems, and are not significantly different from the unweighted mean values.
The best-fit dispersion is equivalent to $\sigma_g = 1.30$ magnitudes.

\begin{figure}[t]
\vspace{-0.1cm}
\begin{center}
\includegraphics[width=1.1\hsize]{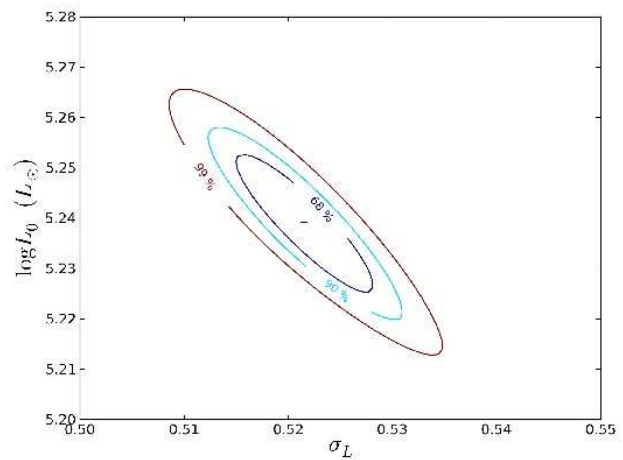}
\end{center}
\vspace{-0.3cm}
\caption{\cfont Confidence intervals of best-fit parameters $L_0$ and $\sigma_L$ of a log-normal fit for the 
global GCLF, with all galaxies combined (Equation~\ref{eq:gaussianfit}).}
\vspace{0.4cm}
  \label{fig:lf_contours}
\end{figure}

Figure~\ref{fig:lf_contours} shows the confidence intervals of the global fit parameters.
As was pointed out long ago by \citet{hanes_whittaker1987}, the two parameters of the fit are partially correlated
if (as is the case here)
the data do not reach clearly fainter than the turnover point (the true $L_0$).  
Thus, for example, an overestimated $\sigma$ leads to a fainter estimated $L_0$.
Nevertheless, the
extremely large statistical size of our database allows us to determine the parameters fairly
precisely.

In the second last column of Table 3, we quantify how well this global GCLF matches each individual 
system by evaluating the goodness of fit 
$\chi^2_{\rm g}$ (without solving for the parameters separately, but simply adopting log $L_0$ = 5.24
and $\sigma = 0.52$).  The fit is calculated in the range of luminosity between 
$L_{0,\rm g}$ and each galaxy's $\Lmax$.  We find that the quality of the global fit is scarcely
worse than the fits optimized to each individual galaxy.
Lastly, we also perform a Kolmogorov-Smirnov (KS) test and 
calculate the probability $P_{KS,g}$ (calculated over the same range as $\chi^2_{\rm g}$ and listed in the last column) 
that the observed cluster samples were drawn from the same \emph{global} distribution. 
Table~\ref{tab:par} shows that at least six systems have $P_{\rm KS,g} > 1\%$, which means they are 
not inconsistent with the universal GCLF.  We note of course that these $P_{\rm KS,g}-$values cannot be as large as similar ones 
calculated one-by-one (where we would be testing only the hypothesis that each individual LF
adequately matches the log-normal model).
The KS probability is low for one galaxy (NGC 6166), primarily because of the somewhat steeper downturn
of the LF at high luminosities.

In summary, we find that \emph{our data for these seven galaxies are consistent with 
a ``universal'' lognormal GCLF shape for BCG-type systems.}
Visually the global fit is indistinguishable from the individual fits. Red lines on 
Figure~\ref{fig:lf_all} overplot the global fit on the histograms of observed cluster numbers, 
adjusting only the normalization to the actual size of the GC system. 

\subsection{Comparisons with Other Galaxies}

Several comparisons can be made for GCLF fits in other giant ellipticals.  \citet{peng_etal2009}, using the same
$F814W$ bandpass for extremely deep HST data around M87, find log $L_0 = 5.06 \pm 0.02$ and 
$\sigma_I = 0.55 \pm 0.02$ dex with an excellent fit to a log-normal model (see their Figure 8).
\citet{harris_etal09} studied the GCLFs of the Coma cluster galaxies and for a composite LF of five gE's
found $\sigma_{L} = 0.59$ dex, log $L_0 \simeq 5.0$ (transformed from the $V-$band), and again
a good match to the log-normal model.  These Coma data do not have as large a total GC sample and do not reach
quite as faint in absolute magnitude as our present BCG sample, so the correlation noted 
above between the turnover point and dispersion may partly explain why the fitted $L_0$ is a bit fainter and
$\sigma_L$ a bit broader than we find.  

\begin{figure*}[ht]
\vspace{-0.1cm}
\begin{center}
\includegraphics[width=0.5\hsize]{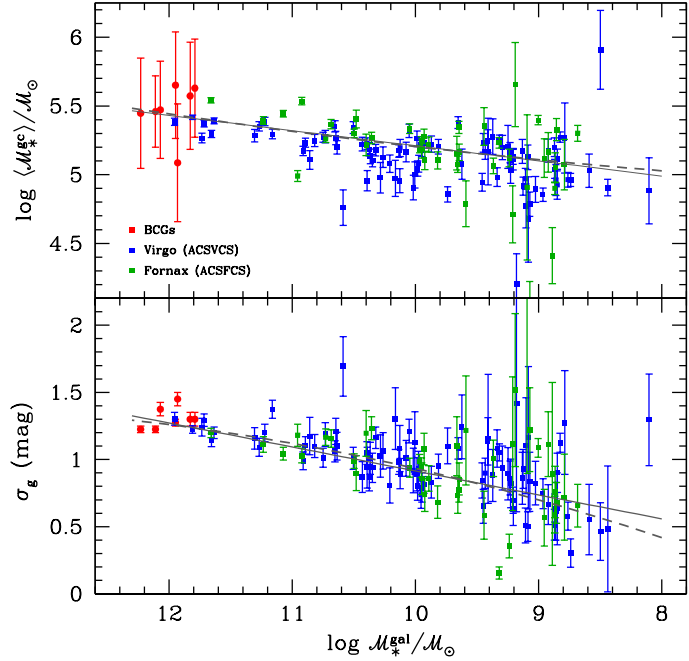}
\end{center}
\vspace{-0.3cm}
\caption{\cfont \emph{Upper panel:} Correlation of GCLF turnover mass with the stellar mass of the host
	galaxy; here, our BCG (red points) data are added to previous measurements for 
	the Virgo (blue points) and Fornax (green points) galaxies
	\citep{villegas2010}.  
	\emph{Lower panel:}  Correlation of the GCLF dispersion with galaxy stellar mass.
	In both panels, the weighted best-fit lines are shown for linear and quadratic
correlations (see text).}
\vspace{0.2cm}
  \label{fig:gclfparams}
\end{figure*}

The 7 BCGs studied here can be added to the trends of GCLF turnover and dispersion with host galaxy luminosity, 
as defined from the many lower-luminosity galaxies in Virgo and Fornax \citep{villegas2010}.  
The results are shown in Figure~\ref{fig:gclfparams}.  In the top panel, $M_{\star}^{gc}$ is
the GC mass corresponding to the GCLF turnover luminosity, calculated with a stellar mass-to-light ratio
from the \citet{into_portinari2013} relations.  For the Virgo and Fornax galaxies imaged with HST/ACS, the
$(g-z)$ color index was used, whereas for the 7 BCGs the color index is $(B-I)$.  
The errorbars in $M_{\star}^{gc}$ for the BCGs (for which only the bright half of the GCLF is measured) are
inevitably larger than for the much nearer Virgo and Fornax members (for which almost the entire run of the
GCLF was measured and the turnover is more precise).  The trend derived by \citet{villegas2010} is
$\Delta \sigma / \Delta M_z = -0.10 \pm 0.01$;
it is clear that the BCGs continue this trend smoothly upward.

The lower panel of Fig.~\ref{fig:gclfparams} shows the same trend for the GCLF dispersion.  Both parameters are plotted
against galaxy stellar mass $M_{\star}^{gal}$, again derived from the \citet{into_portinari2013} color-(M/L)
relations.  Again, the BCGs continue upward along very much the same trend defined by the smaller galaxies.
In both panels, weighted best-fit lines are shown for an assumed linear relation (solid line) and
a quadratic relation (dashed line); these are scarcely different, and in either case indicate that
the BCGs belong to the same family as other, smaller galaxies.

The GCLF dispersion $\sigma$ has also been measured for various giant galaxies at distances $\sim 100$ Mpc
and beyond from SBF (surface brightness fluctuation) analysis.  Though this method is not strictly 
comparable to fully resolved photometry of GCs, the results are consistent with the data listed above.
\citet{blakeslee_etal1997}, from a ground-based imaging survey of BCGs, determined
$\langle \sigma_L \rangle = 0.57$ dex ($1.42 \pm 0.02$ mag) for
14 such galaxies with low internal uncertainties on $\sigma$.  Overall, an empirical picture emerges
in which the intrinsic width of the GCLF (for large galaxies) is consistently in the
range $\sigma_L = 0.52 - 0.55$ dex.

\subsection{Internal Gradients with Galactocentric Distance}

\begin{figure*}[ht]
\vspace{-0.1cm}
\begin{center}
\includegraphics[width=0.4\hsize]{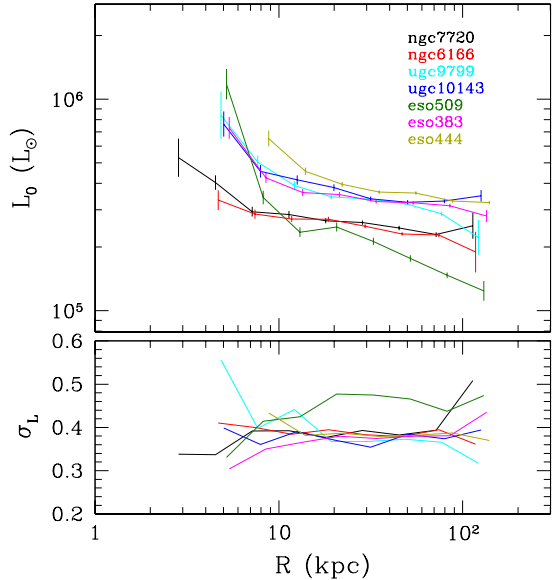}
\end{center}
\vspace{-0.3cm}
\caption{\cfont \emph{Upper panel:} Correlation of the fitted GCLF turnover luminosity $L_0$ with projected galactocentric
	distance $R$.  For $R\gtrsim 20$ kpc, $L_0$ scales roughly as $R^{-0.2}$ in all seven galaxies.
	\emph{Lower panel:}  Correlation of the fitted GCLF dispersion $\sigma_L$ with projected galactocentric distance.}
\vspace{0.2cm}
  \label{fig:lfgrad}
\end{figure*}

We have also briefly investigated how the GCLF might change with projected galactocentric
distance $R$.  In Figure~\ref{fig:lfgrad}, the best-fit solutions for turnover luminosity $L_0$ and dispersion
$\sigma_L$ are plotted against $R$ in kiloparsecs.  For each galaxy the $R-$range was broken into $8-9$
rather narrow zones, a subdivision that was permitted by the very large statistical sample sizes (the relative
numbers of clusters in each zone are indicated by the errorbar sizes in the upper panel of the figure).
We note that although formal solutions were carried out for the regions $R \lesssim 20$ kpc, these should
be given little weight:  
in these inner zones the numbers of clusters per bin are $\lesssim 100$ with few faint GCs, and the gradually
increasing background light from the central galaxy begins to affect the limiting magnitude of the
photometry particularly in the $F814W$ band.

In the range $R \gtrsim 20$ kpc the limiting magnitudes remain uniform (see discussion above) and the
$(L_0, \sigma_L$) fits are more reliable.  In all seven galaxies a consistent pattern for $L_0$
to decrease weakly with $R$ is evident, with an estimated power-law slope $L_0 \sim R^{-0.2}$.  
The best-fit dispersion, however, remains nearly constant with $R$ at $\sigma_L \simeq 0.4$.  
For comparison, the ``global fit'' for the dispersion, summing over all radii, is the somewhat larger value of $\sigma_L \simeq 0.5$.
Given that $L_0$ shows a shallow dependence on $R$, the
solutions within very restricted radial zones are expected to give smaller $\sigma$ than the global average.

Regarding GCLF dependence on halo location, not much exists in the previous literature as a basis
for comparison.  In most cases
the GCLF has been quoted as a single solution for the GC population over the entire galaxy,
or measured over a relatively small radial range determined by detector size so that systematic
changes in $R$ are not easy to see.  In addition, few GCLF studies penetrate outward to the very large galactocentric
distances that we have studied here.  However, for the Milky Way, 
the analysis of \citet{harris01} (see particularly
Fig.~40 and Table 9 there) does show a shallow outward decrease of the GCLF turnover luminosity for $R \gtrsim 6$ kpc,
with a very similar dependence $L_0 \sim R^{-0.2}$ to that found for our BCGs.
By contrast, \citet{tamura_etal06} and \citet{jordan2007} find little change in the turnover luminosity
out to $R \sim 40$ kpc for the Virgo giant M87. \citet{bassino_etal2008} similarly find no strong change
in $L_0$ over $8 - 60$ kpc for the Antlia giants NGC 3258 and 3268.

The simplest interpretation of a gradient in $L_0$ is a similarly shallow trend in mean GC mass, 
but $L_0$ is also influenced by other factors such as metallicity and dynamical evolution.
The latter possibility seems unlikely, since dynamical evolution processes should be relatively ineffective
at such enormous distances in the halo at reshaping the GCLF \citep{gnedin_ostriker97,fall_zhang01,vesperini2001}.
However, the shallow downward gradient may at least partly be explained as a byproduct of a metallicity
gradient in the GC system.  That is, in many large galaxies the mean GC metallicity decreases outward
\citep[e.g.][]{geisler_etal1996,rhode_zepf2004,bassino2006,harris09a,harris09b,liu_etal2011}.  
When we couple this trend with the observation that the 
GCLF turnover also becomes fainter in the smaller galaxies 
in which almost all the GCs are metal-poor
\citep[][]{jordan2007,villegas2010}, then we would expect
to find a net decrease in $L_0$ going further outward into the progressively more metal-poor halo (and if this
argument is correct, a steeper metallicity gradient in the GC system should go along with a steeper gradient in
$L_0$).
But for a fixed mean GC turnover \emph{mass}, the $I-$band \emph{luminosity} of the turnover is almost independent
of metallicity \citep[see the models of][]{ashman_etal1995}.  A possible implication of the trend
we see is therefore that the metal-richer GCs should be systematically more massive than the metal-poor ones.
These issues will be discussed more completely in a later paper containing the metallicity
distribution function measurements more explicitly.

As a final point in this section, we can ask how valid is the Gaussian LF model in the first place.  This 
simple (two-parameter) and extremely well known functional form has been used as a quick and convenient 
descriptor of the GCLF for many decades, but it does not clearly 
result from any underlying physical theory for the
formation and evolution of a GCS.  Other simple forms have been tried, the most notable of which is
probably the ``evolved Schechter function'' introduced by \citet{jordan2007} to fit the GCLFs in the
Virgo galaxies.  \citet{harris_etal09} applied it to the Coma giants as well.  This function is asymmetric,
and thus is preferable as a match for the \emph{entire} luminosity range of GCLF because 
the GCLF is observed to decline more steeply on the faint side of the
turnover $L_0$ than on the much more easily observable bright side.  It also makes a logical link
to the dynamical evolution of the GCS where an assumed initial power-law LF changes into the present-day
form by preferential destruction of the lower-mass clusters 
\citep[e.g. see the discussions of][]{vesperini2001,jordan2007}.
However, for the great majority of cases in the literature 
\citep[see][for a catalog]{harris_etal13} the range $L > L_0$ is the only part covered by the observations,
and over this high-luminosity range both functions do well at matching the empirical data
\citep[see][for a specific comparison in the Coma cluster galaxies]{harris_etal09}.  
More importantly for the present discussion, either functional form leads to our major conclusion
that the present-day GCLFs in BCG systems have a near-universal shape.

\section{Superluminous Clusters?}

In addition to regular clusters well described by the log-normal function, the LFs and CMDs show a
scattering of objects with luminosities of several million $L_{\odot}$ and
above (we note here that {\it very} red objects are excluded by the initial color cuts).  
They are shown by shaded histograms at the upper tail of the distributions in Figure~\ref{fig:lf_all}.
Specifically, we adopt a working definition of superluminous objects as those with luminosities 
above the drop-off limit, $L > \Lmax$.
We can then assess quantitatively how likely these clusters are to appear in the sample if the true GCLF is 
given by our log-normal fit. 

The number of superluminous clusters expected from the log-normal fit is 10, by definition of Equation~(\ref{eq:lmax}).  
This will be the case \emph{if} the LF shape continues to be strictly log-normal to arbitrarily large luminosity.
But in practice, we find that only one system (NGC 6166) contains the expected number, 
while the other six contain ``too many'' superluminous clusters 
at the $> 2 \sigma$ level if $N_{SL}$ is governed simply by Poisson count statistics \emph{and}
if the steep dropoff in $N(L)$ at the high end follows the Gaussian-like model.
The highest number is 29, in ESO383-G076.  These numbers are listed as $N_{SL}$ in Table 3.

As seen directly from the color-magnitude diagrams and the LFs in the previous figures,
these objects extend up to $\sim 10^{8}\Lsun$.  They \emph{may} be the analogs of ultracompact 
dwarf (UCD) galaxies, discovered in the nearby Virgo, Fornax, Hydra, Coma, and Centaurus 
galaxy clusters \citep[e.g.][]{misgeld_etal11,mieske_etal12}.  Such objects are defined as 
having $L \gtrsim 10^{6}\Lsun$ but otherwise are compact enough and metal-poor enough 
to bear some resemblances to globular clusters.  
As noted above, the only limits we can place on their scale sizes are that they have
effective radii $\lesssim 20$ pc, a level that would rule out only the largest UCDs.

UCDs are typically located outside of the stellar effective radius of the host galaxy. We have examined 
the spatial distribution of superluminous clusters and found them to be consistent with the overall location 
of the regular ($L < L_{max}$) clusters.  The projected median distance from the host galaxy center is on the average 
only 20\% larger for the superluminous clusters, and ranges between 2 and $4\kpc$, 
solidly within the locus of the regular GC population.  

\begin{figure}[t]
\vspace{-0.3cm}
\begin{center}
\includegraphics[width=1.1\hsize]{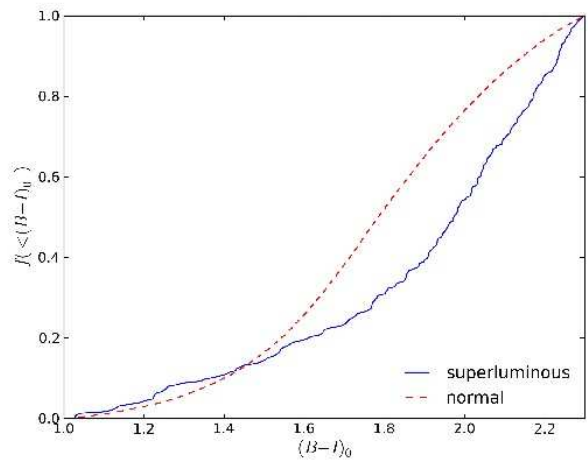}
\end{center}
\vspace{-0.3cm}
\caption{\cfont Cumulative distribution of the $(B-I)_0$ color of regular ($\log{L}<6.7$, dashed line) 
and superluminous ($\log{L}>6.7$, solid line) clusters for the combined sample of all seven systems. 
Superluminous clusters are systematically redder.}
\vspace{0.3cm}
  \label{fig:lf_color}
\end{figure}

We have also compared the color distributions of the regular and superluminous clusters for the 
combined sample of all seven systems. For this purpose, we adopt the lowest common luminosity threshold of 
$\log{\Lmax} = 6.7$ to define the superluminous clusters, and to account for the vast difference 
in relative numbers of the two sets, we consider the cumulative fractional distributions. 
Figure~\ref{fig:lf_color} shows that the superluminous clusters are skewed systematically 
redder than the regular clusters; a KS test indicates that the difference between the two curves is 
significant well above the 99\% level.  In other words, 
the superluminous clusters tend to populate the region above the GC red sequence, though with 
noticeable scatter in color.
Notably, the same sort of red-sequence extension was also seen in the Hydra I BCG, NGC 3311
\citep{wehner_etal08} and in the Coma BCG NGC 4874, but interestingly, \emph{not} in the 
other Coma supergiant NGC 4889 \citep{harris_etal09}.
The brighter UCDs are also typically redder than globular clusters \citep{misgeld_etal11},
arguing that at least part of this superluminous GC population may be identifiable as UCDs.

It is only in these very large BCGs, with enormous numbers of globular clusters to draw from, that we can 
begin seeing these small numbers of superluminous objects as forming a rough ``sequence'' extending upward
from the bulk of the GC population \citep{misgeld_etal11}. Unfortunately, gaining more information about the physical properties 
of the superluminous objects will be difficult, at least for the galaxies studied here.
Their numbers even with this dataset are inevitably small, they are too faint ($I \sim 23, B \sim 25$) 
for easy spectroscopy, and spatial resolutions of $0.03''$ or less will be needed to resolve their
structures with any confidence.  Thus at present our case for identification of superluminous GCs with
UCDs is tentative.

\section{Summary}

We have presented the first results from a new HST imaging survey of the extremely rich globular cluster systems 
around BCG galaxies.  The strongest conclusion from our study is
that these central supergiant galaxies have GC populations that follow a remarkably similar luminosity
distribution that is virtually indistinguishable from the standard log-normal shape, at least for luminosities
higher than the GCLF turnover.

If we adopt a mass-to-light ratio $M/L_{I} \approx 2$ for the old stellar populations of GCs at an
average color $(B-I)_0 \approx 1.7$, according to the empirical relations of \citet{bell_etal03}, then the 
GCLF that we observe translates into a GCMF (mass function) with the 
peak at $M \approx 3\times 10^{5}\Msun$.  The GCLF shape is strongly inconsistent with a single power-law 
from $10^{5}\Msun$ to $10^{7}\Msun$, contrary to what might be
expected from the massive young star clusters seen in present-day merger or starburst environments.
We suggest that if these massive clusters were little affected by dynamical disruption, then (at least at the high-mass end)
the most massive GCs must have formed with an initial mass distribution that was not a single power-law but
rather more nearly resembling a log-normal shape from the start.

In 6 of our 7 target galaxies we also find evidence for an excess of ``superluminous'' clusters
particularly in the color range of the redder, more metal-rich GCs.
Their luminosities and colors (though not their spatial distribution) provide some evidence that
they may be identified as UCDs, though the case cannot be considered to be conclusive as yet.

\section*{Acknowledgements}

Based on observations made with the NASA/ESA Hubble Space Telescope, obtained at the Space Telescope Science Institute, 
which is operated by the Association of Universities for Research in Astronomy, Inc., under NASA contract NAS 5-26555. 
WEH acknowledges financial support from NSERC (Natural Sciences and Engineering Research Council of Canada).
OG was supported in part by NASA and by  
the hospitality of the Aspen Center for Physics, Kavli Institute for Cosmological Physics in Chicago, and
Kavli Institute for Theoretical Physics in Santa Barbara.
DG gratefully acknowledges support from the Chilean BASAL Centro de
Excelencia en Astrof\'isica y Tecnolog\'ias Afines (CATA) grant PFB-06/2007.  BCW acknowledges support from NASA grant 
HST-GO-12238.001-A. 

{\it Facilities:} \facility{HST (ACS)}

\makeatletter\@chicagotrue\makeatother

\bibliographystyle{apj}
\bibliography{gc}

\begin{thebibliography}{67}
\expandafter\ifx\csname natexlab\endcsname\relax\def\natexlab#1{#1}\fi

\bibitem[{{Alamo-Mart{\'{\i}}nez} {et~al.}(2013){Alamo-Mart{\'{\i}}nez},
  {Blakeslee}, {Jee}, {C{\^o}t{\'e}}, {Ferrarese},
  {Gonz{\'a}lez-L{\'o}pezlira}, {Jord{\'a}n}, {Meurer}, {Peng}, \&
  {West}}]{alamo-martinez_etal2013}
{Alamo-Mart{\'{\i}}nez}, K.~A., {Blakeslee}, J.~P., {Jee}, M.~J.,
  {C{\^o}t{\'e}}, P., {Ferrarese}, L., {Gonz{\'a}lez-L{\'o}pezlira}, R.~A.,
  {Jord{\'a}n}, A., {Meurer}, G.~R., {Peng}, E.~W., \& {West}, M.~J. 2013,
  \apj, 775, 20

\bibitem[{{Anderson} \& {Bedin}(2010)}]{anderson_bedin2010}
{Anderson}, J. \& {Bedin}, L.~R. 2010, \pasp, 122, 1035

\bibitem[{{Ashman} {et~al.}(1995){Ashman}, {Conti}, \&
  {Zepf}}]{ashman_etal1995}
{Ashman}, K.~M., {Conti}, A., \& {Zepf}, S.~E. 1995, \aj, 110, 1164

\bibitem[{{Bassino} {et~al.}(2006){Bassino}, {Faifer}, {Forte}, {Dirsch},
  {Richtler}, {Geisler}, \& {Schuberth}}]{bassino2006}
{Bassino}, L.~P., {Faifer}, F.~R., {Forte}, J.~C., {Dirsch}, B., {Richtler},
  T., {Geisler}, D., \& {Schuberth}, Y. 2006, \aap, 451, 789

\bibitem[{{Bassino} {et~al.}(2008){Bassino}, {Richtler}, \&
  {Dirsch}}]{bassino_etal2008}
{Bassino}, L.~P., {Richtler}, T., \& {Dirsch}, B. 2008, \mnras, 386, 1145

\bibitem[{{Baumgardt} \& {Makino}(2003)}]{baumgardt_makino2003}
{Baumgardt}, H. \& {Makino}, J. 2003, \mnras, 340, 227

\bibitem[{{Beasley} {et~al.}(2002){Beasley}, {Baugh}, {Forbes}, {Sharples}, \&
  {Frenk}}]{beasley2002}
{Beasley}, M.~A., {Baugh}, C.~M., {Forbes}, D.~A., {Sharples}, R.~M., \&
  {Frenk}, C.~S. 2002, \mnras, 333, 383

\bibitem[{{Bell} {et~al.}(2003){Bell}, {McIntosh}, {Katz}, \&
  {Weinberg}}]{bell_etal03}
{Bell}, E.~F., {McIntosh}, D.~H., {Katz}, N., \& {Weinberg}, M.~D. 2003, \apjs,
  149, 289

\bibitem[{{Bertin} \& {Arnouts}(1996)}]{bertin_arnouts96}
{Bertin}, E. \& {Arnouts}, S. 1996, \aaps, 117, 393

\bibitem[{{Blakeslee} {et~al.}(1997){Blakeslee}, {Tonry}, \&
  {Metzger}}]{blakeslee_etal1997}
{Blakeslee}, J.~P., {Tonry}, J.~L., \& {Metzger}, M.~R. 1997, \aj, 114, 482

\bibitem[{{Bournaud} {et~al.}(2008){Bournaud}, {Duc}, \&
  {Emsellem}}]{bournaud2008}
{Bournaud}, F., {Duc}, P.-A., \& {Emsellem}, E. 2008, \mnras, 389, L8

\bibitem[{{Bridges} {et~al.}(1996){Bridges}, {Carter}, {Harris}, \&
  {Pritchet}}]{bridges_etal1996}
{Bridges}, T.~J., {Carter}, D., {Harris}, W.~E., \& {Pritchet}, C.~J. 1996,
  \mnras, 281, 1290

\bibitem[{{Brodie} {et~al.}(2011){Brodie}, {Romanowsky}, {Strader}, \&
  {Forbes}}]{brodie2011}
{Brodie}, J.~P., {Romanowsky}, A.~J., {Strader}, J., \& {Forbes}, D.~A. 2011,
  \aj, 142, 199

\bibitem[{{Brodie} \& {Strader}(2006)}]{brodie_strader06}
{Brodie}, J.~P. \& {Strader}, J. 2006, \araa, 44, 193

\bibitem[{{Bromm} \& {Clarke}(2002)}]{bromm_clarke2002}
{Bromm}, V. \& {Clarke}, C.~J. 2002, \apjl, 566, L1

\bibitem[{{Burgarella} {et~al.}(2001){Burgarella}, {Kissler-Patig}, \&
  {Buat}}]{burgarella2001}
{Burgarella}, D., {Kissler-Patig}, M., \& {Buat}, V. 2001, \aj, 121, 2647

\bibitem[{{Chandar} {et~al.}(2011){Chandar}, {Whitmore}, {Calzetti}, {Di Nino},
  {Kennicutt}, {Regan}, \& {Schinnerer}}]{chandar_etal11}
{Chandar}, R., {Whitmore}, B.~C., {Calzetti}, D., {Di Nino}, D., {Kennicutt},
  R.~C., {Regan}, M., \& {Schinnerer}, E. 2011, \apj, 727, 88

\bibitem[{{Elmegreen} {et~al.}(2012){Elmegreen}, {Malhotra}, \&
  {Rhoads}}]{elmegreen2012}
{Elmegreen}, B.~G., {Malhotra}, S., \& {Rhoads}, J. 2012, \apj, 757, 9

\bibitem[{{Fall} \& {Zhang}(2001{\natexlab{a}})}]{fall_zhang2001}
{Fall}, S.~M. \& {Zhang}, Q. 2001{\natexlab{a}}, \apj, 561, 751

\bibitem[{{Fall} \& {Zhang}(2001{\natexlab{b}})}]{fall_zhang01}
---. 2001{\natexlab{b}}, \apj, 561, 751

\bibitem[{{Geisler} {et~al.}(1996){Geisler}, {Lee}, \&
  {Kim}}]{geisler_etal1996}
{Geisler}, D., {Lee}, M.~G., \& {Kim}, E. 1996, \aj, 111, 1529

\bibitem[{{Gieles}(2009)}]{gieles09}
{Gieles}, M. 2009, \mnras, 394, 2113

\bibitem[{{Gieles} \& {Baumgardt}(2008)}]{gieles_baumgardt08}
{Gieles}, M. \& {Baumgardt}, H. 2008, \mnras, 389, L28

\bibitem[{{Gnedin} \& {Ostriker}(1997)}]{gnedin_ostriker97}
{Gnedin}, O.~Y. \& {Ostriker}, J.~P. 1997, \apj, 474, 223

\bibitem[{{Griffen} {et~al.}(2010){Griffen}, {Drinkwater}, {Thomas}, {Helly},
  \& {Pimbblet}}]{griffen2010}
{Griffen}, B.~F., {Drinkwater}, M.~J., {Thomas}, P.~A., {Helly}, J.~C., \&
  {Pimbblet}, K.~A. 2010, \mnras, 405, 375

\bibitem[{{Ha{\c s}egan} {et~al.}(2005){Ha{\c s}egan}, {Jord{\'a}n},
  {C{\^o}t{\'e}}, {Djorgovski}, {McLaughlin}, {Blakeslee}, {Mei}, {West},
  {Peng}, {Ferrarese}, {Milosavljevi{\'c}}, {Tonry}, \&
  {Merritt}}]{hasegan2005}
{Ha{\c s}egan}, M., {Jord{\'a}n}, A., {C{\^o}t{\'e}}, P., {Djorgovski}, S.~G.,
  {McLaughlin}, D.~E., {Blakeslee}, J.~P., {Mei}, S., {West}, M.~J., {Peng},
  E.~W., {Ferrarese}, L., {Milosavljevi{\'c}}, M., {Tonry}, J.~L., \&
  {Merritt}, D. 2005, \apj, 627, 203

\bibitem[{{Hanes} \& {Whittaker}(1987)}]{hanes_whittaker1987}
{Hanes}, D.~A. \& {Whittaker}, D.~G. 1987, \aj, 94, 906

\bibitem[{{Harris}(1996)}]{harris96}
{Harris}, W.~E. 1996, \aj, 112, 1487

\bibitem[{{Harris}(2001{\natexlab{a}})}]{harris2001}
{Harris}, W.~E. 2001{\natexlab{a}}, in Saas-Fee Advanced Course 28: Star
  Clusters, ed. L.~{Labhardt} \& B.~{Binggeli}, 223

\bibitem[{{Harris}(2001{\natexlab{b}})}]{harris01}
{Harris}, W.~E. 2001{\natexlab{b}}, in Star Clusters, Saas-Fee Advanced Course
  28. Lecture Notes 1998, Swiss Society for Astrophysics and Astronomy, ed. by
  L. Labhardt and B. Binggeli (Berlin:Springer), 223--408

\bibitem[{{Harris}(2009{\natexlab{a}})}]{harris09a}
---. 2009{\natexlab{a}}, \apj, 699, 254

\bibitem[{{Harris}(2009{\natexlab{b}})}]{harris09b}
---. 2009{\natexlab{b}}, \apj, 703, 939

\bibitem[{{Harris}(2010)}]{harris2010}
---. 2010, Royal Society of London Philosophical Transactions Series A, 368,
  889

\bibitem[{{Harris} {et~al.}(2013){Harris}, {Harris}, \&
  {Alessi}}]{harris_etal13}
{Harris}, W.~E., {Harris}, G.~L.~H., \& {Alessi}, M. 2013, \apj, 772, 82

\bibitem[{{Harris} {et~al.}(2009){Harris}, {Kavelaars}, {Hanes}, {Pritchet}, \&
  {Baum}}]{harris_etal09}
{Harris}, W.~E., {Kavelaars}, J.~J., {Hanes}, D.~A., {Pritchet}, C.~J., \&
  {Baum}, W.~A. 2009, \aj, 137, 3314

\bibitem[{{Harris} {et~al.}(1995){Harris}, {Pritchet}, \&
  {McClure}}]{harris_etal1995}
{Harris}, W.~E., {Pritchet}, C.~J., \& {McClure}, R.~D. 1995, \apj, 441, 120

\bibitem[{{Harris} \& {Pudritz}(1994)}]{harris_pudritz1994}
{Harris}, W.~E. \& {Pudritz}, R.~E. 1994, \apj, 429, 177

\bibitem[{{Harris} {et~al.}(2006){Harris}, {Whitmore}, {Karakla}, {Oko{\'n}},
  {Baum}, {Hanes}, \& {Kavelaars}}]{harris_etal06}
{Harris}, W.~E., {Whitmore}, B.~C., {Karakla}, D., {Oko{\'n}}, W., {Baum},
  W.~A., {Hanes}, D.~A., \& {Kavelaars}, J.~J. 2006, \apj, 636, 90

\bibitem[{{Howard} {et~al.}(2014){Howard}, {Pudritz}, \& {Harris}}]{howard2014}
{Howard}, C.~S., {Pudritz}, R.~E., \& {Harris}, W.~E. 2014, \mnras, 438, 1305

\bibitem[{{Into} \& {Portinari}(2013)}]{into_portinari2013}
{Into}, T. \& {Portinari}, L. 2013, \mnras, 430, 2715

\bibitem[{{Jord{\'a}n} {et~al.}(2007){Jord{\'a}n}, {McLaughlin},
  {C{\^o}t{\'e}}, {Ferrarese}, {Peng}, {Mei}, {Villegas}, {Merritt}, {Tonry},
  \& {West}}]{jordan2007}
{Jord{\'a}n}, A., {McLaughlin}, D.~E., {C{\^o}t{\'e}}, P., {Ferrarese}, L.,
  {Peng}, E.~W., {Mei}, S., {Villegas}, D., {Merritt}, D., {Tonry}, J.~L., \&
  {West}, M.~J. 2007, \apjs, 171, 101

\bibitem[{{Kravtsov} \& {Gnedin}(2005)}]{kravtsov_gnedin05}
{Kravtsov}, A.~V. \& {Gnedin}, O.~Y. 2005, \apj, 623, 650

\bibitem[{{Kruijssen} {et~al.}(2011){Kruijssen}, {Pelupessy}, {Lamers},
  {Portegies Zwart}, \& {Icke}}]{kruijssen_etal2011}
{Kruijssen}, J.~M.~D., {Pelupessy}, F.~I., {Lamers}, H.~J.~G.~L.~M., {Portegies
  Zwart}, S.~F., \& {Icke}, V. 2011, \mnras, 414, 1339

\bibitem[{{Larsen}(2009)}]{larsen09}
{Larsen}, S.~S. 2009, \aap, 494, 539

\bibitem[{{Li} \& {Gnedin}(2014)}]{li_gnedin14}
{Li}, H. \& {Gnedin}, O.~Y. 2014, \apj, submitted, arXiv:1405.0763

\bibitem[{{Liu} {et~al.}(2011){Liu}, {Peng}, {Jord{\'a}n}, {Ferrarese},
  {Blakeslee}, {C{\^o}t{\'e}}, \& {Mei}}]{liu_etal2011}
{Liu}, C., {Peng}, E.~W., {Jord{\'a}n}, A., {Ferrarese}, L., {Blakeslee},
  J.~P., {C{\^o}t{\'e}}, P., \& {Mei}, S. 2011, \apj, 728, 116

\bibitem[{{Mashchenko} {et~al.}(2008){Mashchenko}, {Wadsley}, \&
  {Couchman}}]{mashchenko2008}
{Mashchenko}, S., {Wadsley}, J., \& {Couchman}, H.~M.~P. 2008, Science, 319,
  174

\bibitem[{{McLaughlin} \& {Fall}(2008)}]{mclaughlin_fall2008}
{McLaughlin}, D.~E. \& {Fall}, S.~M. 2008, \apj, 679, 1272

\bibitem[{{Mieske} {et~al.}(2012){Mieske}, {Hilker}, \&
  {Misgeld}}]{mieske_etal12}
{Mieske}, S., {Hilker}, M., \& {Misgeld}, I. 2012, \aap, 537, A3

\bibitem[{{Mieske} {et~al.}(2010){Mieske}, {Jord{\'a}n}, {C{\^o}t{\'e}},
  {Peng}, {Ferrarese}, {Blakeslee}, {Mei}, {Baumgardt}, {Tonry}, {Infante}, \&
  {West}}]{mieske2010}
{Mieske}, S., {Jord{\'a}n}, A., {C{\^o}t{\'e}}, P., {Peng}, E.~W., {Ferrarese},
  L., {Blakeslee}, J.~P., {Mei}, S., {Baumgardt}, H., {Tonry}, J.~L.,
  {Infante}, L., \& {West}, M.~J. 2010, \apj, 710, 1672

\bibitem[{{Misgeld} {et~al.}(2011){Misgeld}, {Mieske}, {Hilker}, {Richtler},
  {Georgiev}, \& {Schuberth}}]{misgeld_etal11}
{Misgeld}, I., {Mieske}, S., {Hilker}, M., {Richtler}, T., {Georgiev}, I.~Y.,
  \& {Schuberth}, Y. 2011, \aap, 531, A4

\bibitem[{{Muratov} \& {Gnedin}(2010)}]{muratov_gnedin10}
{Muratov}, A.~L. \& {Gnedin}, O.~Y. 2010, \apj, 718, 1266

\bibitem[{{Peng} {et~al.}(2011){Peng}, {Ferguson}, {Goudfrooij}, {Hammer},
  {Lucey}, {Marzke}, {Puzia}, {Carter}, {Balcells}, {Bridges}, {Chiboucas},
  {del Burgo}, {Graham}, {Guzm{\'a}n}, {Hudson}, {Matkovi{\'c}}, {Merritt},
  {Miller}, {Mouhcine}, {Phillipps}, {Sharples}, {Smith}, {Tully}, \& {Verdoes
  Kleijn}}]{peng_etal11}
{Peng}, E.~W., {Ferguson}, H.~C., {Goudfrooij}, P., {Hammer}, D., {Lucey},
  J.~R., {Marzke}, R.~O., {Puzia}, T.~H., {Carter}, D., {Balcells}, M.,
  {Bridges}, T., {Chiboucas}, K., {del Burgo}, C., {Graham}, A.~W.,
  {Guzm{\'a}n}, R., {Hudson}, M.~J., {Matkovi{\'c}}, A., {Merritt}, D.,
  {Miller}, B.~W., {Mouhcine}, M., {Phillipps}, S., {Sharples}, R., {Smith},
  R.~J., {Tully}, B., \& {Verdoes Kleijn}, G. 2011, \apj, 730, 23

\bibitem[{{Peng} {et~al.}(2009){Peng}, {Jord{\'a}n}, {Blakeslee}, {Mieske},
  {C{\^o}t{\'e}}, {Ferrarese}, {Harris}, {Madrid}, \& {Meurer}}]{peng_etal2009}
{Peng}, E.~W., {Jord{\'a}n}, A., {Blakeslee}, J.~P., {Mieske}, S.,
  {C{\^o}t{\'e}}, P., {Ferrarese}, L., {Harris}, W.~E., {Madrid}, J.~P., \&
  {Meurer}, G.~R. 2009, \apj, 703, 42

\bibitem[{{Peng} {et~al.}(2008){Peng}, {Jord{\'a}n}, {C{\^o}t{\'e}},
  {Takamiya}, {West}, {Blakeslee}, {Chen}, {Ferrarese}, {Mei}, {Tonry}, \&
  {West}}]{peng_etal08}
{Peng}, E.~W., {Jord{\'a}n}, A., {C{\^o}t{\'e}}, P., {Takamiya}, M., {West},
  M.~J., {Blakeslee}, J.~P., {Chen}, C.-W., {Ferrarese}, L., {Mei}, S.,
  {Tonry}, J.~L., \& {West}, A.~A. 2008, \apj, 681, 197

\bibitem[{{Penny} {et~al.}(2014){Penny}, {Forbes}, {Strader}, {Usher},
  {Brodie}, \& {Romanowsky}}]{penny2014}
{Penny}, S.~J., {Forbes}, D.~A., {Strader}, J., {Usher}, C., {Brodie}, J.~P.,
  \& {Romanowsky}, A.~J. 2014, \mnras, 439, 3808

\bibitem[{{Rejkuba}(2012)}]{rejkuba2012}
{Rejkuba}, M. 2012, \apss, 341, 195

\bibitem[{{Rhode} \& {Zepf}(2004)}]{rhode_zepf2004}
{Rhode}, K.~L. \& {Zepf}, S.~E. 2004, \aj, 127, 302

\bibitem[{{Saha} {et~al.}(2011){Saha}, {Shaw}, {Claver}, \&
  {Dolphin}}]{saha_etal11}
{Saha}, A., {Shaw}, R.~A., {Claver}, J.~A., \& {Dolphin}, A.~E. 2011, \pasp,
  123, 481

\bibitem[{{Shapiro} {et~al.}(2010){Shapiro}, {Genzel}, \& {F{\"o}rster
  Schreiber}}]{shapiro_etal2010}
{Shapiro}, K.~L., {Genzel}, R., \& {F{\"o}rster Schreiber}, N.~M. 2010, \mnras,
  403, L36

\bibitem[{{Stetson}(1987)}]{stetson87}
{Stetson}, P.~B. 1987, \pasp, 99, 191

\bibitem[{{Tamura} {et~al.}(2006){Tamura}, {Sharples}, {Arimoto}, {Onodera},
  {Ohta}, \& {Yamada}}]{tamura_etal06}
{Tamura}, N., {Sharples}, R.~M., {Arimoto}, N., {Onodera}, M., {Ohta}, K., \&
  {Yamada}, Y. 2006, \mnras, 373, 601

\bibitem[{{Vesperini}(2001)}]{vesperini2001}
{Vesperini}, E. 2001, \mnras, 322, 247

\bibitem[{{Villegas} {et~al.}(2010){Villegas}, {Jord{\'a}n}, {Peng},
  {Blakeslee}, {C{\^o}t{\'e}}, {Ferrarese}, {Kissler-Patig}, {Mei}, {Infante},
  {Tonry}, \& {West}}]{villegas2010}
{Villegas}, D., {Jord{\'a}n}, A., {Peng}, E.~W., {Blakeslee}, J.~P.,
  {C{\^o}t{\'e}}, P., {Ferrarese}, L., {Kissler-Patig}, M., {Mei}, S.,
  {Infante}, L., {Tonry}, J.~L., \& {West}, M.~J. 2010, \apj, 717, 603

\bibitem[{{Wehner} {et~al.}(2008){Wehner}, {Harris}, {Whitmore}, {Rothberg}, \&
  {Woodley}}]{wehner_etal08}
{Wehner}, E.~M.~H., {Harris}, W.~E., {Whitmore}, B.~C., {Rothberg}, B., \&
  {Woodley}, K.~A. 2008, \apj, 681, 1233

\bibitem[{{Whitmore} {et~al.}(2010){Whitmore}, {Chandar}, {Schweizer},
  {Rothberg}, {Leitherer}, {Rieke}, {Rieke}, {Blair}, {Mengel}, \&
  {Alonso-Herrero}}]{whitmore_etal2010}
{Whitmore}, B.~C., {Chandar}, R., {Schweizer}, F., {Rothberg}, B., {Leitherer},
  C., {Rieke}, M., {Rieke}, G., {Blair}, W.~P., {Mengel}, S., \&
  {Alonso-Herrero}, A. 2010, \aj, 140, 75

\bibitem[{{Zhang} \& {Fall}(1999)}]{zhang_fall99}
{Zhang}, Q. \& {Fall}, S.~M. 1999, \apjl, 527, L81

\end{thebibliography}

\label{lastpage}

\end{document}